# Considerations on the P ≠ NP question.

Alfredo von Reckow.

**Abstract**

In order to prove that the P class of problems is different to the NP class, we consider the satisfability problem of propositional calculus formulae, which is an NP-complete problem [1], [3]. It is shown that, for every search algorithm A, there is a set E(A) containing propositional calculus formulae, each of which requires the algorithm A to take a non-polynomial time to find the truth-values of its propositional letters satisfying it. Moreover, E(A) 's size is an exponential function of n, which makes it impossible to detect such formulae in a polynomial time. Hence, the satisfability problem does not have a polynomial complexity.

## 1. Some definitions.

Consider L, the propositional calculus formal theory, as defined in Mendelson [3] and let $L^+$ be an extension of L containing the 0-letter formulas : $\square$ (empty formula, or contradiction) and $\blacksquare$ (tautology). Let $\mathcal{F}$ be the countable set of formulas in $L^+$. Let $\equiv$ denote the "logic equivalence" of formulas, defined by $A \equiv B$ iff A and B yield the same result when evaluating its truth tables.

It can be easily verified that $\equiv$ is a equivalence relationship. Let the quotient set be
$\mathcal{G} = \mathcal{F} / \equiv\ = \{[A] / A \in \mathcal{F}\ \}$.

We will use indistinctly the terms "atom" or "positive literal" for a propositional letter. A "negative literal" is an atom's negation. A "literal" is an atom or the negation of an atom.

### A formula´s cuasinorm.

For each $A \in \mathcal{F}$, we define a quasi-norm $\|A\|$ = number of different atoms in A. This quasi-norm has the following properties:

- For each $A \in \mathcal{F}$, $\|A\| \geq 0$.
- $\|A\| = 0$ iff A is $\square$ or $\blacksquare$.
- Triangular inequality: For A and B in $\mathcal{F}$, and any binary connective op,
  $\|A\ op\ B\| \leq \|A\| + \|B\|$.
  If A and B share atoms, the inequality is strict (including the case when A and B are the same formula). The equality occurs when A and B do not share any atoms (including the case when A or B is $\square$ or $\blacksquare$).

A property like $\|\lambda A\| = |\lambda| \|A\|$ (for $\lambda$ real) is not valid here. One way to define "multiplication" of a formula by a positive integer is $2A = A\ op\ A$, for some binary connective op. But this produces $\|n A\| = A$ for any natural n. Furthermore, the product $\lambda A$ makes no sense for any number $\lambda$ which is not a positive integer.

Since $\|\lambda A\| \neq |\lambda| \|A\|$ for $\lambda$ real, the function $\|\ \|$ is not a norm, so we call it a quasi-norm.

## Irreducibility

Two formulas can be equivalent, despite having different number of atoms. It is possible to have $A \equiv B$ and $\|A\| \neq \|B\|$, because of the absorption laws:

$p \vee (p \wedge q) \equiv p$
$p \wedge (p \vee q) \equiv p$

We will say that a formula A is irreducible of order k iff $\|A\| = k$ and it is not equivalent to another formula B which has less atoms than A. (i.e. if $A \equiv B$ and $\|A\| = k$ implies $\|B\| \geq k$).

In other words, $A \in \mathcal{I}$ is irreducible of order k iff
$\|A\| = k$ and
$(\forall B \in \mathcal{I})(A \equiv B \Rightarrow \|B\| \geq k)$

## A class' cuasinorm.

For $X \in \mathcal{G}$, let's define $\|X\| = \min_{B \in X} \|B\|$, which counts the minimum number of atoms figuring in a formula from the X equivalence class.

Here again, $\|\ \|$ is a quasi-norm, having the following properties:

- For every $X \in \mathcal{G}$, $\|X\| \geq 0$.
- $\|X\| = 0$ iff X is the class of all contradictions (which are equivalent to $\square$) or the class of all tautologies (which are equivalent to ■).
- Triangular inequality: For X and Y in $\mathcal{G}$, and any binary connective op,
 $\|X \text{ op } Y\| \leq \|X\| + \|Y\|$, where we define $X \text{ op } Y = \bigcup \{F_1 \text{ op } F_2\}$ for $F_1 \in X$ and $F_2 \in Y$.

The set of order n irreducible formulas is defined by

$\mathcal{I}_n = \{ A\ /\ \|A\| = \|[A]\| = n\ \}$.

Then we have $\mathcal{I} = \bigcup_{n \in N} \bigcup_{X \in F_n} [X]$,

And the set of all irreducible formulas is $\mathcal{I}_\infty = \bigcup_{n \in N} \mathcal{I}_n$

A formula is simply "irreducible" if it is irreducible of order n, for some natural number n.

We define the convertion function,

$f: \mathcal{I}/\equiv\ \to\ \mathcal{I}_\infty$
 $f([A]) = B$, where $B \equiv A$ and A is irreducible.

from each equivalence class [A], f takes an irreducible formula B. Equivalently, f can be seen as an algorithm (explicity stated) which produces an irreducible formula B from the class [A]. The function f converts any formula in the class [A] to the irreducible formula B.

We have B ∈ $\mathcal{F}_n$, for some natural number n, and [ B ] ∈ $\mathcal{F}_n/\equiv$

We define the normalization function h : $\mathcal{F} \to \mathcal{F}$ , where h(A) is in disjunctive normal form (dnf). Then, for each equivalence class [A], h(f(A)) gets an irreducible and normalized formula.

For any given n, let´s consider the set of "minimal" formulas having n propositional letters $p_1, p_2, \ldots p_n$, which are the order n irreducible formulas.

Let $\mathcal{G}_n$ = { [$A_n$] / $A_n$ is irreducible having order n }. Then $\mathcal{G} = \bigcup_{n \in N} \mathcal{G}_n$

Let $\mathcal{I}_n = \mathcal{F}_n/\equiv$ = { [$A_n$] / B ∈ [$A_n$] iff B is order n irreducible and B ≡ $A_n$ }

Each class of $\mathcal{I}_n$ contains only order n irreducible formulas.

Then $\mathcal{I} = \bigcup_{n \in N} \mathcal{I}_n$ is the set of all equivalence classes having only irreducible formulas.

If we take one element of each equivalence class, we get the set of all irreducible formulas, pair wise non equivalent.

Now we construct the image of the function f, taking the order n minimal formulas:

$\mathcal{H}_n$ = { $B_n$ / $B_n$ = f(X) for X ∈ $\mathcal{I}_n$ } where the $B_n$ are not equivalent to each other, and

$\mathcal{H}_\infty = \bigcup_{n \in N} \mathcal{H}_n.$

Then $\mathcal{H}_\infty$ is the set of all irreducible und unique formulas. In other words, each of the formulas in $\mathcal{H}_\infty$ is irreducible (because it has a minimum number of propositional letters, and it is impossible to eliminate a letter using absorption) and there are no two formulas in $\mathcal{H}_\infty$ being equivalent to each other (which makes each formula unique).

From now on, we will work only with formulas from $\mathcal{H}_\infty$. That is, we will assume that any given formula A is order n irreducible and we will ignore other formulas equivalent to A.

**2. Why this is a search problem?.**

We will show that there is no algorithm to produce directly, in a straightforward way, the truth values for the propositional letters in a formula, which make the formula true, unless there is a search involved. To this end, we show that this satisfability problem is equivalent to construct a "closed" formula for the roots of a polynomial with n variables.

Let's define the function h, from $L^+$ to the set of n-variable polynomials, by

- For a formula A of $L^+$, having n propositional variables $p_1, p_2, \ldots, p_n$, let's associate to each of its letters a numerical variable $x_i$: $h(p_i) = x_i$ for $1 \leq i \leq n$.

- $h(\neg A) = 1 - h(A)$

If A and B are two formulas in $L^+$, then

- $h(A \wedge B) = h(A)\, h(B)$
- $h(A \vee B) = h(A) + h(B) - h(A)\, h(B)$
- $h(A \Rightarrow B) = (1 - h(A))(1 - h(B)) + h(B)$
- $h(A \Leftrightarrow B) = (1 - h(A))(1 - h(B))(1 + h(A) + h(B)) + h(A)\, h(B)$

Note that

$h(A)^n$ is equivalent to $h(A)$ in the following sense:
$h(A)^n = 0$ iff $h(A) = 0$ and $h(A)^n = 1$ iff $h(A) = 1$

Note that h is a truth function assigning to a formula A a value 0 or 1 when each of the variables $x_i$ takes a value of 0 or 1.

In particular, in the resulting polynomial we can replace any factor $x^n$ by x.

For example, for the idempotency law: $A \wedge A \equiv A$ we have:

$h(A \wedge A) = h(A)\, h(A) = h(A)^2 \equiv h(A)$,

where the last "$\equiv$" means equivalence in the sense that we get the same values for $h(A)^2$ and for $h(A)$ when each of the corresponding variables $x_i$ take a value of 0 or 1.

Similarly, with the other idempotency law, $A \vee A \equiv A$, we get
$h(A \vee A) = h(A) + h(A) - h(A)h(A) = 2h(A) - h(A)^2 \equiv 2h(A) - h(A) = h(A)$.

**Tautologies and contradictions:**

If a formula A is a tautology, then $h(A) \equiv 1$.
If a formula A is a contradiction, then $h(A) \equiv 0$.

Example 1.

For the tautology $A \Rightarrow A$ we have:
$h(A \Rightarrow A) = (1 - h(A))^2 + h(A) \equiv (1 - h(A)) + h(A) = 1$.

This latter equivalence can be proved in the following ways:

a) $(1 - h(A))^2 \equiv (1 - h(A))$, because both sides have the same value when $h(A) = 0$ or when $h(A) = 1$.

b) $(1 - h(A))^2 + h(A) = 1 - 2h(A) + h(A)^2 + h(A) = 1 - h(A) + h(A)^2$
$\equiv 1 - h(A) + h(A) = 1$

Example 2.

For the contradiction $A \wedge \neg A$ whe have:

$h(A \wedge \neg A) = h(A)(1 - h(A)) = h(A) - h(A)^2 \equiv h(A) - h(A) = 0$.

For any formula A, the problem of finding the truth-values for its variables $p_1, p_2, \ldots, p_n$ satisfaying A (i.e. making it true) is equivalent to the problem of finding the corresponding numerical values for n-tuple $(x_1, x_2, \ldots, x_n)$ which are zeros of the function g, where $g(A) = h(A) - 1$.

This function is a polynomial with the n variables $(x_1, x_2, \ldots, x_n)$, which will be called the formula's "characteristic polynomial", and has the form $g(A) = q$, where

$q : [0, 1]^n \to R$, takes values on the n-dimentional unit cube, and is defined by

(1) $$q(x_1,\ldots,x_n) = \sum_{i=1}^{r(n)} \alpha_i x_1^{\beta_{i1}} x_2^{\beta_{i2}} \ldots x_n^{\beta_{in}} + \gamma$$

where each exponent $\beta_{ki}$ has a value of 0 or 1, the coefficients $\alpha_i$ and $\gamma$ are integers, and the variables $x_i$ have a value of 0 or 1, and $r(n) = 2^n$.

We define $0^0 = 1$.

The problems then is to find a value for each variable $x_i$, either a 0 or a 1, in such a way that $q(x_1, x_2, \ldots, x_n) = 0$. Or equivalently:

(2)  Find the roots of the polynomial q
(3)  With the restrictions $x_i$ in $\{0, 1\}$ for $i = 1, \ldots, n$.

In other words, find $q^{-1}(0) \cap \{0, 1\}^n$.

**Lemma.**

There is no closed formula to calculate directly the roots of this type of polynomials. Hence, there is no algorithm to find directly (without making any search) the truth values of the propositional letters satisfying a formula A.

Proof.

The equation $g(A) = 0$ has n variables. It can be seen as a "system" of 1 equation with n unknowns. This is an under determined system, having infinite real roots. To determine them, it suffices to solve g(A) = 0 for $x_1$, which will then depend on the other independent variables $x_2$, ..., $x_n$.

Since these variables are independent (there is no relationship between them), each of them can take any real value. Once we chose a value for each variable $x_2$, ..., $x_n$, that will produce a value for $x_1$.

From these infinite set of values for the n-tuple $(x_1, ..., x_n)$ .we have to separate those where all the variables are 0 or 1.

Because there are no additional conditions on the variables, no further relationship between them can be obtained, and this happens just because the variables $x_2$, ..., $x_n$ are independent. There is only the restriction (3) on the value of the solutions, but it is of no use for finding a closed formula for the solutions of (2).

Hence, we have no other way left, but to carry on an exhaustive search, and we have no additional information derived from the formula which could lead us to take any shorcuts and reduce the number of cases (or number of rows in the truth table) to be considered.

**Example.**

For the formula

$A : (p_1 \lor p_2 \lor p_3) \land \neg (p_1 \land p_2) \land \neg (p_1 \land p_3) \land \neg (p_2 \land p_3)$

The associated polynomial for A is

$h(A) = x_1 + x_2 + x_3 - 2x_1x_2 - 2x_1x_3 - 2x_2x_3 + 3x_1x_2x_3$

The characteristic polynomial is

$g(A) = x_1 + x_2 + x_3 - 2x_1x_2 - 2x_1x_3 - 2x_2x_3 + 3x_1x_2x_3 - 1$

and the equation g(A) – 1 = 0 is

$x_1 + x_2 + x_3 - 2x_1x_2 - 2x_1x_3 - 2x_2x_3 + 3x_1x_2x_3 - 1 = 0$.

With the restrictions $x_i \in \{0, 1\}$ for i = 1,.., 3.

In this equation, the variable $x_1$ depends on $x_2$ and $x_3$. This two last variables have no further relationship to one another. If $x_3$ takes any given value, it has no influence at all on the values that $x_2$ can take. The only thing we know is that we are interested in the cases where $x_2$ as well as $x_3$ take a value of 0 or 1.

Hence, we have no choice but to make an exhaustive search, using the 4 possible values for the duple $(x_2, x_3)$. The solutions will be those cases where the equation makes sense and yielding values of 0 or 1 for $x_1$.

Note that if $(x_2, x_3) = (1, 1)$ then $x_1$ vanishes and the equation becomes $1 = 0$, which makes no sense. Hence $(x_2, x_3) \neq (1, 1)$.

From other point of view, solving for $x_1$ we get

(4) $$x_1 = \frac{2x_2 x_3 - x_2 - x_3 + 1}{3x_2 x_3 - 2(x_2 + x_3) + 1}$$

where the denominator is zero if $(x_2, x_3) = (1, 1)$. It is zero for other values of these variables, but it does not matter, because in none of those other cases both variables have a value of 0 or 1.

Note that $x_2$ and $x_3$ can take any real value in (4), as long as the denominator does not vanish. There are infinite values for $x_2$ and $x_3$ satisfying (4), but we are interested only in the 4 cases where they are 0 or 1, and there are no shorcuts for finding them. We must try all the options and see which ones work.

In the general case, for a formula having n propositional letters, we can not avoid the need to explore $2^{n-1}$ options for the variables $x_2, \ldots x_n$.

Other way to search is the following:

If we put arbitrarily $x_1 = 0$ in (4), then $2x_2 x_3 - x_2 - x_3 + 1 = 0$, and from here we get $x_2 = (x_3 - 1) / (2x_3 - 1)$. Here we have two cases: if si $x_2 = 0$, then $x_3 = 1$ and if $x_2 = 1$, then $x_3 = 0$.

If we put $x_1 = 0$ in (4), then $2x_2 x_3 - x_2 - x_3 + 1 = 3x_2 x_3 - 2(x_2 + x_3) + 1$, so $x_2 = x_3 / (x_3 - 1)$. Here again we have two cases: if $x_2 = 0$, then $x_3 = 0$ and if $x_2 = 1$, then $x_3 = x_3 - 1$, which is impossible.

We have had to analize 4 different cases, for 3 variables, corresponding to $2^{n-1}$ options, for n = 2, and the variables $x_1, \ldots x_{n-1}$

We can solve for any of the n variables as a function of the other n-1 variables. In any event, we must try $2^{n-1}$ options.

**Notes.**

1. Note that the characteristic polynomial g has n variables and has $2^n$ terms. Each term is a product of k variables, and there are $C(n, k) = n! / (k! (n-k)!)$ combinations of terms having k varibles. Adding that up, there are

$$\sum_{k=0}^{n} \binom{n}{k} = 2^n$$

terms, each one has one coefficient and n exponents. Any algorithm processing g has thus an input of $(n+1) 2^n$ numbers, an exponential function of n, which can not be processed in a polinomial time.

Moreover, if there were a "closed form" formula to determine the binary roots of g, then it would depend on the $2^n$ coefficients of g. Since the variables $x_2, \ldots, x_n$ are independent, a closed form for each of them would be a function of those $2^n$ coefficients, and it cannot be calculated in a time $t = q(n)$, where q is a polynomial.

Such a formula, if it existed, would behave not better than a sequential exhaustive search.

2. The characteristic polynomial g has n variables $x_1, x_2, \ldots, x_n$ each one with an exponent 0 or 1. In any occurrence of any of this variables the exponent can be replaced by an odd natural number k, and the resulting polynomial has the same "binary roots" (those being 0 or 1) of g. The exponent must be 0 or odd to avoid de case $(-1)^2 = 1$.

Hence, g is equivalent to infinitely many polynomials (in the sense that all of them have the same binary roots).

If there were an explicit formula to find the binary roots of g, that same formula would yield the binary roots of all the polinomials q equivalent to g, which can have an arbitrarily high degree.

These comes from the idempotency laws $A \wedge A \equiv A$ and $A \vee A \equiv A$.

3. It is not possible to determine, a priori, additional conditions to find the binary roots of g. If there were n-1 such additional conditions, then we would have a non-linear system of n equations with n variables.

The number of solutions for the characteristic polynomial can be anything from 0 (in the case of a contradiction) to $2^n$ (for a tautology). If all this options are possible, then no additional conditions can exist. For a formula like $p_1 \vee p_2 \vee \ldots \vee p_n$, having $2^n - 1$ solutions, any additional condition must be redundant. For a formula having $2^n - 2$ solutions, there can be at most one additional condition, and son on. The number of additional conditions depends on the number of solutions, which is not known in advance, so we don't know beforehand how many conditions are necessary for a given formula.

Moreover, if such conditions depend on the number of solutions, and the solutions depend on the coefficients of the characteristic polynomial, then those conditions will depend on those coefficients. Since there are $2^n$ coefficients in the characteristic polynomial, the additional conditions will depend on $2^n$ inputs, which can not be processed in a polynomial time.

On the other hand, if we introduce, arbitrarily, any condition $g_1(x_1, \ldots, x_n) = 0$, then some roots can be missed.

For example, if we put arbitrarily the condition $x_1 = 0$, then we will miss all the roots where $x_1 = 1$, in the cases where such roots exists. In the cases where g has a factor $(x_1 - 1)$ then we will miss all its binary roots.

If we put a condition like $x_1 - x_2 = 0$, then we might miss the roots where $x_1 \neq x_2$, and we do not know beforehand if that is indeed the case. Simple conditions as these can be, of course, verified beforehand, but more complex conditions can not.

In the example

$A : (p_1 \vee p_2 \vee p_3) \wedge \neg (p_1 \wedge p_2) \wedge \neg (p_1 \wedge p_3) \wedge \neg (p_2 \wedge p_3)$

The characteristic polynomial is

$x_1 + x_2 + x_3 - 2x_1x_2 - 2x_1x_3 - 2x_2x_3 + 3x_1x_2x_3 - 1 = 0$.

If we add the condition $x_1 - x_2 = 0$, (which can be considere a mere supposition) the characteristic polynomial becomes

$x_3 - x_1x_3 - 1 = 0$.

Where $x_1$ must be zero and $x_3$ must be 1. $x_2$ can be 0 or 1.

The triplet $(0, 1, 0)$ is a root of the characteristic polynomial, but the other triplet $(0, 1, 1)$ is not.

Adding an arbitrary condition not only leads us to miss two roots, but it introduces a fake root.

We have no additional information and we cannot add it arbitrarily, because we don't where to start looking for the roots.

As have been noted in the literature (see [3]), if we know the roots (or if we a have a candidate for root, somehow "guessed" or estimated), we can replace it in the characteristic polynomial and check it in a polynomial time.

But if we don't have a root, we have to search for it, in a sequential way. We must try all the options.

4. Multiplying the factors of g to get the explicit form (1) requires $o(m^n)$ multiplications, where m is the number of factors and n is the number of variables. Instead of carrying on the multiplications, we can consider the associated polynomial, when it is already factored and find its roots. The remaining n-tuplets (those not being a binary root of h) will be the binary root of the characteristic polynomial g.

The idea is to take advantage when h is already factored (as this might take less operations).

In the example,
$A : (p_1 \vee p_2 \vee p_3) \wedge \neg (p_1 \wedge p_2) \wedge \neg (p_1 \wedge p_3) \wedge \neg (p_2 \wedge p_3)$

The associated polynomial is

$h(A) = (x_1 + x_2 + x_3 - x_1x_2 - x_1x_3 - x_2x_3 + x_1x_2x_3)(1 - x_1x_2)(1 - x_1x_3)(1 - x_2x_3)$

h(A) = 0 if al least one of its factors is zero.

The last thre factors give the options

$x_1 = 1, x_2 = 1$, any value for $x_3$
$x_1 = 1, x_3 = 1$, any value for $x_2$
$x_2 = 1, x_3 = 1$, any value for $x_1$

The first factor gives

$x_1 = (x_2 + x_3 - x_2 x_3) / (1 - x_2 - x_3 + x_2 x_3) = U / (1 - U)$, for $U \neq 0$,

Where $U = x_2 + x_3 - x_2 x_3$. All the options are:

| $x_2$ | $x_3$ | U | $x_1$ |
| --- | --- | --- | --- |
| 0 | 0 | 0 | 0 |
| 0 | 1 | 1 | not definided |
| 1 | 0 | 1 | not definided |
| 1 | 1 | 1 | not definided |

The roots of the associated polynomial h are

| $x_1$ | $x_2$ | $x_3$ |
| --- | --- | --- |
| 0 | 0 | 0 |
| 0 | 1 | 1 |
| 1 | 0 | 1 |
| 1 | 1 | 0 |
| 1 | 1 | 1 |

Since $g = h - 1$, for any other triplet $(x_1, x_2, x_3)$, $h(x_1, x_2, x_3)$ is not zero, and hence it must be $h(x_1, x_2, x_3) = 1$. Those other three triplets must be the binary roots of g. They are:

| $x_1$ | $x_2$ | $x_3$ |
| --- | --- | --- |
| 0 | 0 | 1 |
| 0 | 1 | 0 |
| 1 | 0 | 0 |

But this has required to explore all the options for the first factor. It takes an $o(2^{n-1})$ number of operations, for $n = 3$.

The cases when two or more factors are simultaneously zero produce repeated roots, that have to be eliminated from the list of roots.

## 3. Exhaustive search

If there is no direct ("closed form") formula to find the binary roots of the characteristic polynomial of a given formula, we must carry on a sequential search, checking a formula's truth table in a sequential way, one row after another. Since there are no additional conditions to guide the search, it must be made sequentially.

For the set of order n irreducible formulas, we choose an order of evaluation for the propositional letters and for the truth value for each atom. First, we take some permutation of the letters. Then, for every letter we choose if it takes the value 0 or 1 first. Then for every order of evaluation there are formulas taking necessarily an exponential time to find an n-tuple of truth values satisfying it.

### 3. 1. Sequential search algorithms.

In this section we consider "uninformed" or "blind" search algorithms, which uses no additional information (other than the given formula itself) to guide the search or to reduce the number of options. As was shown in section 2, such information is not available beforehand, and if it were, taking it into account requires an exponential order algorithm, which is no better than a sequential uninformed search.

For this reason, in this section we well consider only algorithms exploring a formula's truth table in some order that does not depend on the formula. That is, an algorithm A will explore the truth table's rows in the same fashion for all formulas.

A "sequential search algorithm" assigns truth values to the propositional letters $p_1$, $p_2$, …, $p_n$ occurring in a formula F in some pre-established order (out of the n! permutations of the n letters) and for each letter it is established if 0 or 1 is used first (out of the $2^n$ options).

Consider the set of all sequential search algorithms. This set has n! $2^n$ algorithms, because the n letters can be arranged in n! different permutations and each letter can be evaluated to 0 first and then to 1, or the other way around. If a formula F has n atoms ( $\| F \| = n$) there are n! $2^n$ different ways to evaluate F, depending on the order the atoms are evaluated and if the 0 or 1 is evaluated first for every atom.

Two sequential search algorithms $A_1$ and $A_2$ are "sequentially equivalent" if and only if they assign the same truth values to the same atom in the same order. The algorithms can use different functions or procedures, yielding the same results in the same order, but their implementations can yield different running times on a given machine, operating systems and programming language.

Two sequential search algorithms $A_1$ and $A_2$ are not equivalent, and will be considered different, if they take the atoms in a different order or if they assign the truth values 0 and 1 in a different order to the same atom.

Equivalently, two sequential search algorithms $A_1$ and $A_2$ are different if they explore the truth table rows in a different order.

This sequential search is equivalent to a depth-first search in a binary tree, with a given priority for the order in which the atoms are evaluated and the truth value given to each atom.

Sequential equivalence is an equivalence relationship, so we can take an algorithm from each equivalence class. We will assume that we take the algorithm taking the least time in a particular implementation.

In order to find out if a given formula F is satisfacible, and when it is, find the corresponding truth-values of its atoms, a sequential search algorithm A explores the truth table rows in a certain order. The algorithm A "solves" the formula F if it finds a truth-value assignment for F's atoms, with which F is evaluated true.

We will represent the truth table for F, as evaluated by A, by $A(F) = (b(1), b(2), \ldots, b(2^n))$, where $b(k)$ is the value obtained in row k.

We define the function L as the function giving the first row where algorithm A finds a value of 1 when it evaluates the formula F. This is $L(A(F)) = \min_{1 \leq k \leq 2^n} \{ k \,/\, b(K) = 1 \}$.

**Exploring with the initial order.**

Take $\mathcal{I}_n$, the set of order n irreducible formulas. Take a formula F in $\mathcal{I}_n$. Then $\|F\| = n$, and consider $p_1, p_2, \ldots, p_n$ as the n distinct atoms figuring in F, taken in the same order they appear in F, from left to right. We will call this the "initial order" of the atoms in F.

Denote by $A_{i,j}$ the algorithm which uses the i-th permutation of the atoms of F and the j-th way of giving the truth values to them.

The sequential search algoritm $A_{1,1}$ explores F's truth table's rows in the "natural" order, taking the atoms $p_1, p_2, \ldots, p_n$ in the same order they appear in F and for every one of them it evaluates first the 0 and then the 1 values.

Then, for the formula

F:   $\neg p_1 \wedge \neg p_2 \wedge \ldots \wedge \neg p_n$

the sequential search algoritm $A_{1,1}$ must check all the rows in the truth table, to find a solution only in the last row (that is, in row $2^n$ ). Thus, $L(A_{1,1}(F)) = 2^n$.

**Changing 0s and 1s.**

Take now all the $A_{1,j}$ algorithms. All of them evaluate the atoms $p_1, p_2, \ldots, p_n$ in the order 1, 2, …, n. They differ from each other because they assign the values 0 and 1 in a different order for every letter (one of them will assign first 0 and later 1 to some atom $p_i$, while other one assigns first 1 and later 0 to the same letter $p_i$ ).

More precisely, if the base 2 representation of j is $j = d_1 d_2 \ldots d_n$, then the atom $p_k$ takes first the truth value $d_k$ and later it will take the truth value $1 - d_k$, for $1 \leq k \leq n$.

For every j, if the $A_{1,j}$ algorithm evaluates first true the atom $p_k$ (that is, if $d_k = 1$), then let`s take the letter $q_k$ equal to $p_k$. Otherwise, take $q_k$ equal to $\neg p_k$. Then the formula

(5)     F:     $\neg q_1 \wedge \neg q_2 \wedge \ldots \wedge \neg q_n$

is true only when all the $q_j$ are false. To get there, the algorithm $A_{1,j}$ must explore first all the other rows in the truth table. In this case, we also have Thus, $L(A_{1,j}(F)) = 2^n$.

Thus, the algorithm $A_{1,j}(F)$ has an order $o(2^n)$ temporal complexity.

Permuting the order of the atoms in F is irrelevant, since is has no effect on the rows that have to be evaluated. In any case, the formula F will take an exponential time: $L(A_{i,j}(F)) = 2^n$, for $1 \leq i \leq n!$

**The set $C_n(A_{i,j})$.**

For an integer n given, we define the set $C_n(A_{i,j})$ as the set of order n irreducible and pair-wise non equivalent formulas for whom algorithm $A_{i,j}$ takes a non polynomial time.

The formula F given above in (5) is in $C_n(A_{i,j})$. It the number of formulas in this set (representing each one an equivalence class) where a polynomial depending on n, we could detect this kind of formulas before applying the algorithm. Then we $A_{i,j}$ would execute in a polynomial time for the other formulas in $\mathcal{F}_n$.

This is not the case, as it will be shown that $C_n(A_{i,j}) = o(g(n))$, where g is an exponential-type function.

For every sequential search algorithm $A_{i,j}$ and every formula F, it is posible to apply a pre-process to determine if F is equivalent to a worst-posible case formula, (and in this case, to evaluate it at once). Let's call this revised algorithm $A_{i,j}(1)$.

But then there is another formula $F(1)$ forcing the revised algorithm $A_{i,j}(1)$ to search all the way thru the penultimate row in the truth table of $F(1)$, because that is the only row where $F(1)$ is true. If we repeat the process to construct algorithm $A_{i,j}(2)$, which detects F and $F(1)$, then the formula $F(2)$, which is true only in the next-to-penultimate row, will take an exhaustive search from row 1 to row $2^{n-2}$, and so on.

More precisely, we have two steps in the analysis:

**Lemma 1.**

**For every natural number m, $1 \leq m \leq 2^n$, there is a formula F having n atoms, that is true only in row m.**

For each sequential search algorithm $A_{i,j}$ and every $m \in \{1, \ldots, 2^n\}$, we construct a formula being true only on row m, by taking the formula

F:     $\neg pi_1 \wedge \neg pi_2 \wedge \ldots \wedge \neg pi_n$

which is true only in the last row of the truth table, and changing each atom for its negation, depending on the order in which algorithm $A_{i,j}$ assigns values to each atom.

To this end, we take the base 2 representation of the index j: $j = d_1 d_2 \ldots d_n$. Suppose the atom $p_k$ takes first the value $d_k$ and later on the value $1 - d_k$.

Take the base 2 representation of m: $m = e_1 e_2 \ldots e_n$,

If $e_k = d_k$, then $q_{ik}(m)$ is $p_{ik}$, otherwise, $q_{ik}(m)$ is $\neg p_{ik}$.

Then the formula $G_{ij}(m)$: $\quad \neg q_{i1}(m) \wedge \neg q_{i2}(m) \wedge \ldots \neg q_{in}(m)$

Is true only in row m and it is false in all the other rows. $\square$

**Lemma 2. (from the end of the truth table).**

For every sequential search algorithm $A_{i,j}$, there is an order n irreducible formula F, that is true only in the last 2 rows explored by the algorithm.

**Proof.**

This formula is constructed taking the disjunction

$G_{ij}(2^n-1) \vee G_{ij}(2^n)$
For this formula, algorithm $A_{ij}(1)$ must explore $2^n-1$ rows in the truth table.

Of course, we can detect this formulae and give them a different treatment, with a pre-process.

Call $A_{ij}(2)$ be the algorithm obtained by adding such pre-process to $A_{ij}(1)$, to eliminate the "worst possible cases" for $A_{ij}$ and $A_{ij}(1)$.

Consider now the following formula, which is true only in the last three rows, as explored by algorithm $A_{ij}(2)$:

$G_{ij}(2^n-2) \vee G_{ij}(2^n-1) \vee G_{ij}(2^n)$

As well as the formulas

$G_{ij}(2^n-2) \vee G_{ij}(2^n-1)$
$G_{ij}(2^n-2) \vee G_{ij}(2^n)$

We can separate this formulas with a new pre-process, to get the algoritm $A_{ij}(3)$, and so on.

Continuing this way, for each algorithm $A_{ij}$ we construct a succession of algorithms $A_{ij}(1), A_{ij}(2), A_{ij}(3), \ldots$. Every one of them has a set of worst-possible case formulas, and every one avoids the worst-possible case for the previous algorithms.

The worst-possible case for $A_{ij}(2^{n-1})$ is a formula whose truth table is true in all its second-half rows and false in all its first-half rows. This algorithm has exponential complexity.

To find out if a formula F is true from row k, for $k = 2^n, 2^n-1, 2^n-2, \ldots, 2^{n-1} + 1$, we have to check out $2^{n-1}$ cases. If none of them happens, then we must explore sequentially rows 1 to $2^{n-1}$. The worst possible case occurs with the formula which is true only in row $2^{n-1}$.

The time complexity is therefore:

Rows to check out in the pre-process:         $2^n - (2^{n-1} + 1) + 1 = 2^{n-1}$
Rows to explore in the process                $2^{n-1}$
--------------------------------------------------------
Total                                                       $2^n$

Moreover, for every k, algorithm $A_{ij}(k)$ has exponential complexity.

We have to check first (in the pre-process) for k worst-possible case situations of $A_{ij}(h)$ $h = 1, \ldots, k-1$. For the formulas passing this check, there are the worst-possible cases $A_{ij}(k)$ demanding to explore $2^n - k$ rows. Hence, in the worst-possible case, $A_{ij}(k)$ must explore $k + (2^n - k) = 2^n$ rows.        □

**Conclusion:**

For every way used to explore a truth-table, there is a formula requiring to explore all the other rows first. If we try to avoid this worst-possible cases adding a pre-process to the search algorithm, then there is another formula requiring also to explore all the other rows first.

This applies to any sequential search algorithm that does a "blind" or "uninformed" search, where no additional information is used to guide the search or to discard options, which will be analized in section 3.2.

**Additional notes.**

**1. Additional complexity.**

There are additional sources of complexity: each equivalence class in $\mathcal{G} = \mathcal{F} / \equiv$ contains formulas with an arbitrarily large number of atoms, which are equivalent to a formula with n atoms, due to the absorption laws.

This requires to first simplify a formula, adding extra time to the process. For an n-atom formula F, it can be constructed infinite many other formulas with m atoms, for any $m > n$, equivalent to F by the absorption laws. Because of this, an algorithm processing formulas with n atoms will have to deal also with formulas having $m > n$ atoms to find out if they can be simplified to a formula with n atoms. Since m can be arbitrarily large, the time needed will be arbitrarily large too.

That's why we have supposed that the input formulas for any algorithm are already order-n irreducible.

**2. Extending lemma 2.**

The $A_{ij}$ algorithms take into account only permutations in the order of evaluation of the atoms and of the truth value for each atom. There are $2^n$ n! such permutations. The truth table for an n-order formula has $2^n$ rows, having $2^n$! permutations. Since the proof of lemma 2 includes a constructive method to build a formula that is true in any given row (and only in that row), that proof can be extended to cover the case of an arbitrary permutation of rows of the truth table.

### 3.2. Heuristic Algorithms.

An heuristic algorithm does not explore all the rows in a formula's truth table. It eliminates arbitrarily some of the rows, hoping there is a solution in the remaing rows. Such algorithm will have a polynomial complexity if it explores only $\phi(n)$ rows for an order n irreductible formula F, whre $\phi$ is a polynomial.

This means that the algorithm eliminates a priori all the other rows. The algorithm will be "acceptable" for a formula F if some rows yielding 1 are among the explored rows.

### Lema 3.

For every polynomial complexity heuristic algorithm H, there is a set of formulas C(H) for which H is not acceptable. That is, for every formula in F in C(H), H does not explore any row of F where the result is 1.
In other words, we can not eliminate rows arbitrarily, because that might eliminate precisely those rows having a result of 1.

### Proof.

Let F be an order n irreducible formula and $\phi$ a polynomial depending only on n. Let Hj be one of the $\binom{2^n}{\phi(n)}$ algorithms having polynomial complexity of order $\phi(n)$, which explores only rows 1, 2, ..., $\phi(n)$, evaluating the atoms $p_1$, $p_2$, ..., $p_n$ in the same order they figure in F, from left to right.

For any n-order formula, each algorithm $H_j$ evaluates only rows $i_1$, $i_2$, ..., $i_{\phi(n)}$. This rows are the same for all n-order formulas. The row indexes depend only on the algorithm j-index and not on any given formula.

Take $k_1$ as one of the rows not explored by $H_j$. With the same method used in lemma 2, we can construct a formula that is true only in row $k_1$:

If atom $p_i$ is true in row k then take $q_i$ equal to $p_i$. Otherwise take $q_i$ equal to $\neg p_i$. Take the conjunction of the $q_i$.

$G_{k1}$:     $q_1 \wedge q_2 \wedge \ldots q_n$

This formula is true only in row k, and algorithm $H_j$ missed it, since it does not explore this row.

( Moreover, $H_j$ will miss a solution for any disjunction of formulas $G_{k1} \vee G_{k2} \vee _{...} \vee G_{kr}$, where $k_1$, $k_2$, ..$k_r$ are not explored by $H_j$)

Let $H_j(1)$ be the extensión of $H_j$ obtained by adding a pre-process to check row $k_1$ beforehand. Then we can construct in a similar way another formula where $H_j(1)$ fails for some formula $G_{k2}$, where row $k_2$ is not explored by $H_j(1)$. And then we can extend $H_j(1)$ to $H_j(2)$, wich in turn will fail for some formula $G_{k3}$.

This produces a list of algorithms $H_j$, $H_j(1)$, $H_j(2)$, …, $Hj(t)$ The list is finite, with a restriction: the last index, t, must depend polynomially on n: $t = q(n)$ for some polynomial q.

But the truth table has $2^n$ rows an hence we always can take one excluded row and build a formula being true only on that row, causing $H_j(t)$ to fail, no matter what polynomial function q is used. □

In conclusion, there is a set $C(H_j)$ having a number of equivalence classes w, where w is an exponential function of n, and for any formula F in a class X of $C(H_j)$, $H_j$ does not evaluate the rows where F is true, and therefore misses every solution for F.

**4. Conversion algorithms.**

It is quite simple to find the truth-value assignations for the atoms satisfiying a formula expressed in disjunctive normal form (dnf). There are 3 cases:

1. The formula is a tautology, and is equivalent to ■. It is true for any assignation of truth values to its atoms.

2. The formula is a contradiction, and is equivalent to □. It is false for any assignation of truth values to its atoms. In each of its "disjoint" the formula must have the conjunction of a complementary pair (if some disjoint does not have any such conjunction, this disjoint would have a truth value assignment satisfying the formula, which is impossible).

3. The formula is a contingency. It has at least one "disjoint" and it suffices one of them to be true.

In any case, it can be immediately checked out if the formula is satisfacible or not. If a formula is not equivalent to a contradiction, it must be satisfacible. All we need is to check sequentially the disjoints for the absence of complementary pairs.

This leads to the following algorithm:

To find the truth values satisfying a formula F

1. Convert F to its disjunctive normal form.
2. Check the disjoints sequentially from left to right until a disjoint D having no complementary pairs is found.
3. Assign 1 to each literal in D. This will make some atoms true and some false.

Apparently this is a polynomial complexity algorithm. The problem is: the conversion process (to get a formula's d.n.f.) is a search process in a tree, having exponential complexity.

**Lema 4.**

For every convertion algorithm there is an order n formula F requiring a time t = q(n) where q is an exponential type function.

**Proof.**

A clause is a disjunction of literals, some of them positive and some negative ones. For a given value of n, take a formula F in conjunctive normal form, having k clauses, each having m literals (obviously m ≤ n). Converting F to d.n.f. we get $m^k$ disjoints, where each of them has k literals.

If k = n and m is close to n, the number of disjoints is in the order of $n^n$. □

Note that evaluating the truth table requires to evaluate $2^n$ rows, much less than $n^n$, although both algorithms have exponential complexity.

**5. Asymptotic properties.**

For a given order n, take a representative from each equivalence class $\mathcal{F}_n/\equiv$. We can identify each class with its corresponding truth table. For any n-order formula, its truth table has $2^n$ rows. Since each row can have a resulting 0 or 1, there are $2^{2^n}$ different truth tables, and therefore there are $2^{2^n}$ equivalence classes.

In one half of all classes in $\mathcal{F}_n/\equiv$ the formulas have true in the first row of its truth table. In the other half, the formulas have false in the first row.

The formulas belonging to $2^{2^n-1}$ classes have 1 in its first row, while the formulas belonging to the other $2^{2^n-1}$ classes have a 0 in its first row.

If we put a 0 in row 1, then every one of the remaining $2^n$-1 rows can have a 0 or 1, giving a total of $2^{2^n-1}$ options.

In one fourth of all classes in $\mathcal{F}_n/\equiv$ the formulas have 1 in row 2, knowing they have 0 in row 1. In a similar way, knowing rows 1, 2, …, m-1 to have a 0, there are $2^{2^n-m}$ equivalence classes having a 1 in row m.

The number of classes whose formulas have its first 1 in row 1 or 2 or … or m equals the sum

$$q(m) = 2^{2^n-1} + 2^{2^n-2} + 2^{2^n-3} + ... + 2^{2^n-m} = 2^{2^n}(2^{-1} + 2^{-2} + 2^{-3} + ... + 2^{-m})$$

$$= 2^{2^n}\left(1-\left(\frac{1}{2}\right)^m\right)$$

The percentage of classes whose formulas are true for the first time in one of the rows 1, 2, …, m is then:

$$r(m) = 100\left(1-\left(\frac{1}{2}\right)^m\right) = 100\frac{2^m-1}{2^m}$$

Let π be a s degree polynomial, for a fixed s. If m = π(n), then r(m) is the percentage of classes whose formulas can be solved in a polynomial time (meaning by "solve" to find a n-tuple of truth values satisfying the formula).

Take $\pi(x) = x^s$. Then $m = n^s$ and $r(m) = r(n^s) = 100\left(1-\left(\frac{1}{2}\right)^{n^s}\right)$

We get $\qquad \lim_{n\to\infty} r(n^s) = \lim_{n\to\infty} 100\left(1-\left(\frac{1}{2}\right)^{n^s}\right) = 100\%$.

**Conclusion:**

When n increases, the ratio of equivalence classes having satisfactible formulas, which are detected by a sequential search algorithm exploring only the first π(n) rows (or any set of pre-determined π(n) rows) approaches 100%.

**Notes**.

The ratio of order n satisfactible formulas not solved by a sequential search algorithm exploring π(n) rows decreases to 0 as n increases.

However, the number of equivalence classes increases very fast, as shown in the following table:

|   | Rows in table | Equivalence classes ( = number of Truth tables) |
|---|---|---|
| n | u = $2^n$ | $2^u$ |
| 1 | 2 | 4 |
| 2 | 4 | 16 |
| 3 | 8 | 256 |
| 4 | 16 | 65.536 |
| 5 | 32 | 4294967296 |
| 6 | 64 | 1,84467E+19 |
| 7 | 128 | 3,40282E+38 |
| 8 | 256 | 1,15792E+77 |
| 9 | 512 | 1,3408E+154 |
| 10 | 1.024 | #¡NUM! |
| 11 | 2.048 | #¡NUM! |

| 12 | 4.096 | #¡NUM! |
|---|---|---|
| 13 | 8.192 | #¡NUM! |
| 14 | 16.384 | #¡NUM! |
| 15 | 32.768 | #¡NUM! |
| 16 | 65.536 | #¡NUM! |
| 17 | 131.072 | #¡NUM! |
| 18 | 262.144 | #¡NUM! |
| 19 | 524.288 | #¡NUM! |
| 20 | 1.048.576 | #¡NUM! |

Even tough the ratio of equivalence classes (causing exponential complexity to an algorithm) vs. total number of classes decreases, this latter number increases quite fast, as the following table shows:

| n | 1 | 2 | 3 | 4 | 5 | 6 | 7 | 8 | 9 | 10 |
|---|---|---|---|---|---|---|---|---|---|---|
| 1 | 2 | 2 | 2 | 2 | 2 | 2 | 2 | 2 | 2 | 2 |
| 2 | 4 | 1 | | | | | | | | |
| 3 | 32 | | | | | | | | | |
| 4 | 4096 | 1 | | | | | | | | |
| 5 | 134217728 | 128 | | | | | | | | |
| 6 | 2,8823E+17 | 268435456 | | | | | | | | |
| 7 | 2,65846E+36 | 6,04463E+23 | | | | | | | | |
| 8 | 4,52313E+74 | 6,2771E+57 | | | | | | | | |
| 9 | 2,6187E+151 | 5,5453E+129 | | | | | | | | |
| 10 | 1,7556E+305 | 1,4181E+278 | 16777216 | | | | | | | |
| 11 | #¡NUM! | #¡NUM! | 6,8946E+215 | | | | | | | |
| 12 | #¡NUM! | #¡NUM! | #¡NUM! | | | | | | | |
| 13 | #¡NUM! | #¡NUM! | #¡NUM! | | | | | | | |
| 14 | #¡NUM! | #¡NUM! | #¡NUM! | | | | | | | |
| 15 | #¡NUM! | #¡NUM! | #¡NUM! | | | | | | | |
| 16 | #¡NUM! | #¡NUM! | #¡NUM! | 1 | | | | | | |
| 17 | #¡NUM! | #¡NUM! | #¡NUM! | #¡NUM! | | | | | | |
| 18 | #¡NUM! | #¡NUM! | #¡NUM! | #¡NUM! | | | | | | |
| 19 | #¡NUM! | #¡NUM! | #¡NUM! | #¡NUM! | | | | | | |
| 20 | #¡NUM! | #¡NUM! | #¡NUM! | #¡NUM! | | | | | | |

In the blank entries the calculation formula is not applicable since $2^n < n^s$. The #¡NUM! means an overflow; the numeric limits of Excel on a Pentium computer has been exceeded.

**Increasing speed for the function r.**

As the following table shows, $r(n^s) / 100$ increases quite fast.

| | s | | | | | | | | | |
|---|---|---|---|---|---|---|---|---|---|---|
| n | 1 | 2 | 3 | 4 | 5 | 6 | 7 | 8 | 9 | 10 |
| 1 | 0,5 | 0,5 | 0,5 | 0,5 | 0,5 | 0,5 | 0,5 | 0,5 | 0,5 | 0,5 |
| 2 | 0,75 | 0,9375 | 0,99609375 | 0,999984741 | 1 | 1 | 1 | 1 | 1 | 1 |
| 3 | 0,875 | 0,998046875 | 0,999999993 | 1 | 1 | 1 | 1 | 1 | 1 | 1 |
| 4 | 0,9375 | 0,999984741 | 1 | 1 | 1 | 1 | 1 | 1 | 1 | 1 |
| 5 | 0,96875 | 0,99999997 | 1 | 1 | 1 | 1 | 1 | 1 | 1 | 1 |
| 6 | 0,984375 | 1 | 1 | 1 | 1 | 1 | 1 | 1 | 1 | 1 |
| 7  | 0,9921875    | 1 | 1 | 1 | 1 | 1 | 1 | 1 | 1 | 1 |
| 8  | 0,99609375   | 1 | 1 | 1 | 1 | 1 | 1 | 1 | 1 | 1 |
| 9  | 0,998046875  | 1 | 1 | 1 | 1 | 1 | 1 | 1 | 1 | 1 |
| 10 | 0,999023438  | 1 | 1 | 1 | 1 | 1 | 1 | 1 | 1 | 1 |
| 11 | 0,999511719  | 1 | 1 | 1 | 1 | 1 | 1 | 1 | 1 | 1 |
| 12 | 0,999755859  | 1 | 1 | 1 | 1 | 1 | 1 | 1 | 1 | 1 |
| 13 | 0,99987793   | 1 | 1 | 1 | 1 | 1 | 1 | 1 | 1 | 1 |
| 14 | 0,999938965  | 1 | 1 | 1 | 1 | 1 | 1 | 1 | 1 | 1 |
| 15 | 0,999969482  | 1 | 1 | 1 | 1 | 1 | 1 | 1 | 1 | 1 |
| 16 | 0,999984741  | 1 | 1 | 1 | 1 | 1 | 1 | 1 | 1 | 1 |
| 17 | 0,999992371  | 1 | 1 | 1 | 1 | 1 | 1 | 1 | 1 | 1 |
| 18 | 0,999996185  | 1 | 1 | 1 | 1 | 1 | 1 | 1 | 1 | 1 |
| 19 | 0,999998093  | 1 | 1 | 1 | 1 | 1 | 1 | 1 | 1 | 1 |
| 20 | 0,999999046  | 1 | 1 | 1 | 1 | 1 | 1 | 1 | 1 | 1 |

The number 1 occurs in most places where the fraction subtracted from 1 is too small for the precision handled by Excel.

**Other notes.**

1. In the preceeding análisis, we have not taken into account the process needed to find out if a formula is irreducible, since it would add processing time. A formula may be evaluated with all the its atoms, without trying to simplify it beforehand via the absorption equivalences.

2. If $m < n$ then $r(n) < 100$, because the contradiction class is excluded (its table yields only zeros).

3. It has been shown that for every uninformed sequential search algorithm there is a set of formulas taking an exponential time. An "inverse" problem can be stated as: for every formula there is an uninformed sequential search algorithm giving all the formula's solutions in the first rows of the truth table, and zeros on the remaining rows. The problem is to find the algorithm. This requires a search on the set of uninformed sequential search algorithms. This search has exponential complexity.

If a an n order formula F has m solutions (that is, there are m rows where F is true), then there are $m!(2^n - m)!$ ways to permute the rows keeping the true rows first and the false rows last. The ratio of this number of algorithms to the total ways of exploring F's truth table is $m!(2^n - m)! / (2^n)! = \dfrac{1}{\binom{2^n}{m}}$, which makes it highly unlikely to find one of this algorithms by chance. Using a systematic search, the vast majority of a huge number of algorithms must be tested before a solution can be found.